\documentclass[twocolumn,english,showpacs,preprintnumbers,amsmath,amssymb,floatfix]{revtex4-1}
\usepackage{xcolor}
\usepackage[T1]{fontenc}
\usepackage[latin9]{inputenc}
\usepackage{color}
\usepackage{array}
\usepackage{amstext}
\usepackage{graphicx}
\usepackage{esint}
\usepackage{rotating}
\usepackage{soul}
\usepackage[font=small,labelfont=bf]{caption}
\usepackage[colorlinks = true,
linkcolor = blue,
urlcolor  = blue,
citecolor = blue,
anchorcolor = blue]{hyperref}

\makeatletter

\@ifundefined{textcolor}{}
{%
 \definecolor{BLACK}{gray}{0}
 \definecolor{WHITE}{gray}{1}
 \definecolor{RED}{rgb}{1,0,0}
 \definecolor{GREEN}{rgb}{0,1,0}
 \definecolor{BLUE}{rgb}{0,0,1}
 \definecolor{CYAN}{cmyk}{1,0,0,0}
 \definecolor{MAGENTA}{cmyk}{0,1,0,0}
 \definecolor{YELLOW}{cmyk}{0,0,1,0}
 }

\@ifundefined{definecolor}
 {\usepackage{color}}{}
\@ifundefined{definecolor}
 {\usepackage{color}}{}
\makeatother
\usepackage{babel}

\begin{document}


\title { QCD analysis of non-singlet structure functions at NNLO accuracy, based on the Laplace transform }

\author{S.~Mohammad Moosavi Nejad$^{a,b}$}
\email{Mmoosavi@yazd.ac.ir (Corresponding author)}

\author{Maral Salajegheh$^{a}$}
\email{M.Salajegheh@stu.yazd.ac.ir}

\author{ Abolfazl Mirjalili$^{a}$}
\email{A.Mirjalili@yazd.ac.ir}

\author {S.~Atashbar Tehrani$^{b}$}
\email{Atashbar@ipm.ir}

\affiliation {
$^{(a)}$Physics Department, Yazd University, P.O.Box 89195-741, Yazd, Iran              \\
$^{(b)}$School of Particles and Accelerators, Institute for Research in Fundamental Sciences (IPM), P.O.Box
19395-5531, Tehran, Iran           \\
      }

\date{\today}

%
%
\begin{abstract}\label{abstract}
In this work, using the Laplace
transformation technique we present our results  for non-singlet  quark distributions as well as nucleon structure
function $F_2(x,Q^2)$ in unpolarized case at next-to-next-to-leading order (NNLO) QCD accuracy. We shall particularly compare our results for the  sets of valence-quark parton distribution functions  with the  contemporary collaborations like CT14, CT18, MMHT14, MKAM16 and NNPDF. To construct the nucleon structure function we employ  the expansion of Jacobi polynomials  which  is a suitable transform to convert the results of non-singlet structure function from the Laplace $s$-space to Bjorken $x$-space.
We shall also consider the contributions of target mass correction as well as the higher twist effects at large-$x$ region for the proton and deuteron structure functions. Our results for the  unpolarized quark distribution functions and nucleon structure functions are  in good agreement with recent  theoretical models and available experimental data.
\end{abstract}
%


\maketitle


%
\section{Introduction}
\label{Introduction}

As is well-known, our understanding of Quantum Chromodynamics (QCD) as well as the nucleon structure does profoundly depend on the  deep inelastic scattering (DIS) processes. In this regard,  the theory of QCD  contains  the necessary ingredients  for the  scale
evolution of nucleon structure functions. On this base the proton, as a specific state of nucleons, can be  described in perturbative QCD (pQCD) framework in terms of  parton distribution functions (PDFs)
which are known as the nonperturbative part of  evolution process. The  PDFs  are typically related to the partons, i.e.,  gluon and  quarks,  and refer to the probability to find partons  carrying away a specific fraction of proton's momentum.
In other words,  the nonperturbative inputs  are the PDFs at the initial energy scale  which can be evolved to higher scales within the pQCD framework. Finally, they denote   the
probability density to find a parton which  is carrying a specific fraction of proton's momentum at a desired energy scale.

As usual, these nonperturbative PDFs can be determined by fitting  the available experimental  data including the DIS processes and the ones from hadron colliders~\cite{Ball:2017nwa,Bourrely:2015kla,Harland-Lang:2014zoa,Hou:2017khm,Alekhin:2017kpj,Khanpour:2017slc,Goharipour:2018yov,Soleymaninia:2018uiv,Nejad:2015oca,Nejad:2015far,MoosaviNejad:2012ju,Shoeibi:2017zha}. To get numerical
solutions for the evolved non-singlet PDFs, there are various  methods such as the
Brute-force~\cite{Cabibbo:1978ez,Miyama:1995bd,Hirai:1997gb}, the Laguerre transformation technique~\cite{Toldra:2001yz,Furmanski:1981ja,Blumlein:1989pd,Kumano:1992vd,Kobayashi:1994hy,GhasempourNesheli:2015tva} and the
Mellin-transform~\cite{Gluck:1989ze,Graudenz:1995sk,Blumlein:1997em,Blumlein:2000hw,Stratmann:2001pb},  and for the
analytical solutions we can refer to  the Jacobi polynomials model ~\cite{Khorramian:2008yh,Khorramian:2009xz} as
well as  Laplace transformation technique~\cite{Khanpour:2016uxh,MoosaviNejad:2016ebo,Boroun:2015hiy}.  The evolution of PDFs in $x$-space  can be also done using the common programs like HOPPET \cite{hop}, QCDNUM \cite{qcd}, and APFEL \cite{apf}.

In this paper, using the Laplace transformation technique we compute the non-singlet  quark distributions as well as the unpolarized nucleon structure function  at next-to-next-to-leading order (NNLO) of QCD analysis. To extract the NNLO PDFs  we apply the deep-inelastic data for non-singlet QCD analysis employing the Laplace transformation at  NNLO accuracy. Our results for the  valence-quark PDFs  are also compared  with the  famous Collaborations {{} like CT14, CT18, MMHT14, MKAM16 and NNPDF}. The expansion of Jacobi polynomials  is also employed to construct the nucleon structure function. This is a convenient approach to convert the results of  non-singlet structure function from the Laplace $s$-space to the Bjorken $x$-space.
We will also consider the corrections due to the target mass  as well as the higher twist effect for the proton and deuteron structure functions.  These effects improve {{} the quality of fit for  structure functions} at low energy scales.

The structure of our paper is organized as follows:
In Sec.~\ref{Laplace}, we present an analytical solution for NNLO non-singlet quark density based on the Dokshitzer-Gribov-Lipatov-Altarelli-Parisi (DGLAP) equations in Laplace $s$-space. Sec.~\ref{Jacobi} is devoted to the Jacobi  polynomial technique which yields the non-singlet nucleon structure function in $x$-space.  Our global analysis of valence-quark densities is presented in Sec.~\ref{sec:dataset} where we describe our procedure for the QCD fit of non-singlet $F_2$ structure function data.  Corrections due to the target mass and the Higher twist effect are discussed in Sec.~\ref{Sec:TMC}. Our results and discussions are listed in Sec.~\ref{Sec:Results}.\\
As a supplement to this work, our analytical results  for the Willson coefficient functions as well as  the splitting functions in the
Laplace $s$-space at NLO and NNLO accuracies are presented in appendix $A$.

\section{NNLO Non-singlet solution in Laplace space}
\label{Laplace}

In this work, our main interest is to investigate  the proton spin-independent  structure function, $F_2^p (x, Q^2)$, in non-singlet case at the NNLO QCD
analysis, specifically, in large values of $x$.
Some analytical solutions of the DGLAP equations based on the Laplace transform technique have been recently presented, see for example Refs.~\cite{Block:2010du,Block:2011xb,Block:2010fk,Block:2009en,Block:2010ti,Zarei:2015jvh,Taghavi-Shahri:2016ktz,Boroun:2015cta,Boroun:2014dka}
which contain remarkable success from phenomenological point of view.
In Ref.~\cite{Khanpour:2016uxh}, using the mentioned technique  authors have presented their spin-independent analysis of the structure function  $F_2^p (x, Q^2)$ at NLO accuracy. In Refs.~\cite{MoosaviNejad:2016ebo} and \cite{Sheibani:2018gxt}, the application of Laplace transform technique to the analysis of charged-current structure functions $xF_3(x,Q^2)$ as well as the EMC effects are studied. The spin-dependent structure functions $x g_1^p (x, Q^2)$ are analyzed via the same technique in Refs.~\cite{AtashbarTehrani:2013qea,Salajegheh:2018hfs} at NLO and NNLO
accuracies.

For the non-singlet sector of quark densities, presented by $q_{\rm NS}(x,Q^2)$,  one can write  the following  DGLAP evolution equations at the NNLO expansion
\begin{eqnarray}\label{eq:nonsinglet-DGLAP}
&&\frac{4 \pi}{\alpha_s(Q^2)}  \frac{\partial q_{\text{NS}}}{\partial \ln Q^2} (x,Q^2)  =
q_{\text{NS}} \otimes \Big(p_{NS}^{\text{LO}} + \nonumber  \\
&& \frac{\alpha_s(Q^2)}{4\pi}p_{NS}^{\text{NLO}}
+ (\frac{\alpha_s(Q^2)}{4\pi})^2 p_{NS}^{\text{NNLO}} \Big)(x,Q^2)\;.
\end{eqnarray}

Here, the symbol $\otimes$ denotes the convolution integral and  $\alpha_s(Q^2)$ stands for the renormalized strong coupling constant. Moreover, the
non-singlet Altarelli-Parisi splitting kernels up to three-loops corrections are presented by
$p_{NS}^{LO}(\alpha_s(Q^2))$, $p_{NS}^{NLO}(\alpha_s(Q^2))$ and $p_{NS}^{NNLO}(\alpha_s(Q^2))$,  respectively.
Considering these kernels, the required splitting function has the following expansion
\begin{eqnarray}\label{eq:nonsinglet-Spliting}
p_{NS}(\alpha_s(Q^2)) & = & p_{NS}^{\text{LO}}(x) + \frac{\alpha_s(Q^2)}{4 \pi}p_{NS}^{\text{NLO}} (x)
\nonumber \\
& + &  \left(\frac{\alpha_s(Q^2)}{4 \pi}\right)^2 p_{NS}^{\text{ NNLO} } (x) \,.
\end{eqnarray}

A brief description to extract the analytical solution of the valence quark distribution function through the DGLAP evolution equations based on the
Laplace transform technique is now at hand. Considering the explicit form of  convolution integral in Eq.~(\ref{eq:nonsinglet-DGLAP}), in addition to the $x$-variable an extra $z$-variable would be also appeared.
Taking the variable changes  as $\nu \equiv \ln(1/x)$ and $w \equiv \ln(1/z)$, the  evolution
equation \eqref{eq:nonsinglet-DGLAP} is expressed in terms of the convolution integrals   as
\begin{eqnarray}\label{eq:nonsinglet}
&&	\frac{\partial \hat{F}_{\text{ NS}}}{\partial\tau}(\nu,\tau)   =
\int_0^\nu   \hat{F}_{\text{ NS}}(w,\tau)e^{-(\nu-w)}\,dw \Big(p_{NS}^{\text{ LO}}(\nu-w)   +  \nonumber \\
&& \frac{\alpha_s(\tau)}{4\pi}  p_{NS}^{\text{ NLO}}(\nu-w)+
(\frac{\alpha_s(\tau)}{4\pi})^2
p_{NS}^{\text{ NNLO}}(\nu-w) \Big) \,.
\end{eqnarray}

As is seen, the $q_{\text{NS}}(x,Q^2)$-function in Eq.~(\ref{eq:nonsinglet-DGLAP})  is now presented by $\hat{F}_{\text{NS}}$, including new  variables
$\nu$ and $\tau$. The $\tau$-variable is utilizing to present
the $Q^2$-dependence of Eq.~(\ref{eq:nonsinglet}) and is  entirely given by
$\tau (Q^2, Q_0^2) = (1/4\pi) \int_{Q_0^2}^{Q^2}  \alpha_s (Q'^2)  d\ln Q'^2$.

The Laplace transform on $\hat F_{NS}(\nu,\tau)$ now leads to  $f_{NS}(s,\tau) \equiv  {\cal L} [ \hat F_{NS}(\nu,\tau); s]$. On the other hand, it is
known that the Laplace transform of convolution factors will yield  the ordinary product of the Laplace transform of those factors~\cite{Block:2010du,Block:2011xb,MoosaviNejad:2016ebo}. Therefore, by imposing the Laplace transform on Eq.~\eqref{eq:nonsinglet}, the result
would be  the ordinary first order differential equation  with respect to the $\tau$-variable in Laplace space $s$. The result for the
non-singlet distributions $f_{\rm NS}(s, \tau)$ reads
\begin{eqnarray}\label{eq:nonsinglet-Laplace-space}
 	\frac{\partial f_{\text{ NS}}}{\partial \tau}(s, \tau) && = \left ( \Phi_{\text{ NS}}^{\text{LO}} +
 \frac{\alpha_s(\tau)}{4\pi}\Phi_{\rm NS}^{\text{NLO}}+ (\frac{\alpha_s(\tau)}{4\pi})^2\Phi_{\text{ NS}}^{\text{
 NNLO}} \right )   \nonumber  \\
 	&&\times f_{\text{ NS}}(s,\tau)\,,
\end{eqnarray}
where,  $\Phi_{\text{NS}}^{\text{i}} (\text{i=LO, NLO, NNLO})$ represents the Laplace transform of the splitting kernels at the
desired order. The analytical results for the  splitting kernels in $x$-space are given in Refs.~\cite{Curci:1980uw,Moch:2004pa}. From Eq.~(\ref{eq:nonsinglet-Laplace-space}), one can achieve a  solution involving a simple form as
 \begin{eqnarray}\label{eq:solve-nonsinglet}
f_{\text{ NS}}(s, \tau) = e^{\tau\Phi_{\text{NS}}(s)} \, f^0_{\text{ NS}}(s).
 \end{eqnarray}
In this equation the splitting function, $\Phi_{\text{NS}}(s)$, can be presented in terms of new expansion parameters as
 \begin{eqnarray}\label{eq:fi-nonsinglet}
 	\Phi_{\text{NS}}(s) \equiv \Phi_{\text{ NS}}^{\text{LO}}(s) + \frac{\tau_2}{\tau}\Phi_{\text{NS}}^{\text{
 NLO}}(s)+\frac{\tau_3}{\tau}\Phi_{\text{NS}}^{\text{NNLO}}(s).
 \end{eqnarray}
 
The variables $\tau_2$ and $\tau_3$
in Eq.~\eqref{eq:fi-nonsinglet}  are given by
\begin{equation}
\tau_2 \equiv \frac{1}{(4\pi)^2}\int_{Q_0^2}^{Q^2}\alpha_s^2(Q'^2)d \ln Q'^2\,,
\end{equation}
and
\begin{equation}
\tau_3 \equiv \frac{1}{(4\pi)^3}\int_{Q_0^2}^{Q^2}\alpha_s^3(Q'^2)d \ln Q'^2,
\end{equation}
which are $Q^2$-dependent.\\
For the splitting function $\Phi_{\text{NS}}^{\text{LO}}(s)$ in the Laplace $s$-space, the result is typically given by
\begin{eqnarray}
\Phi_f^{\text{LO} } = 4 - \frac{8}{3}\left(\frac{1}{s + 1} + \frac{1}{s + 2} + 2\left(\gamma_E + \psi (s +
1)\right)\right), \nonumber  \\
\end{eqnarray}
where, $\gamma_E=0.577216\cdots$ is the Euler constant and $\psi(s)=d\ln\Gamma(s)/ds$ is the digamma function. The corresponding results for the  NLO and NNLO  splitting kernels  are too lengthy so they are presented in appendix~{\bf A}.

Considering Eq.~(\ref{eq:solve-nonsinglet}) and using Eq.~(\ref{eq:fi-nonsinglet}) the solution of non-singlet sector of quark densities can be obtained in the Laplace $s$-space. Applying the inverse Laplace transform in two stages one achieves  the quark densities in Bjorken $x$-space as a function of $Q^2$. Details of this procedure can be found in Ref.~\cite{Khanpour:2016uxh}. Back to Eq.~(\ref{eq:solve-nonsinglet}), we see that the solution of non-singlet parton densities would be obtained provided that the $f^0_{\text{ NS}}(s)$ at the initial scale $Q_0$ is known. In this regards, one needs to extract the parton densities at the initial scale $Q_0$ which can be done by a global fit over whole available  data.  One of the reliable  techniques to extract the initial  parton densities is the Jacobi  polynomials approach which will be illustrated in the following sections.

\section{Jacobi polynomials and non-singlet structure function}
\label{Jacobi}

In this section, we apply the Jacobi polynomials to convert the results for the non-singlet structure function from the Laplace $s$-space to the known Bjorken $x$-space. Based on this method,   we will be able to perform a global fit over all available experimental data to extract free parameters in the proposed form of the parton densities at the initial scale $Q_0^2$. This procedure will be explained later. In this approach, the results which will be obtained for the parton densities in  Laplace
$s$-space can be considered as the  moment of densities.
Therefore, the theoretical perspectives on  Jacobi polynomial approach is to extract the non-singlet structure
function $F_2^{NS}(x, Q^2)$ from the NNLO analytical  solution of non-singlet DGLAP equations in Laplace $s$-space. It should be noted  that,  the solution of evolved structure functions at any  value of $x$ and Q$^2$ could be attained through  the method described. This is {{} a crucial feature} for the phenomenological implications.

Now, the results in Laplace $s$-space for the non-singlet structure function $F_2^{NS}(x,Q^2)$  are considered as the Laplace $s$-space moments ${\cal
M}^{NS}(s,Q^2)$. These provide the possibility to include DIS data for performing  a QCD analysis up to NNLO accuracy.
The DIS data contains a wide range of the transferred momentum from  $Q_0^2\gtrsim2GeV^2$ to
$Q^2\sim30000GeV^2$ where the Jacobi polynomials technique works, reasonably.

To construct the moments of  proton structure
function,  ${\cal M}^{p}_2(s, \tau)$, at NNLO accuracy the following combination of parton densities at the valence region $x\geq0.3$  in Laplace $s$-space (for the non-singlet sector) are required
\begin{eqnarray} \label{eq:Fp}
{\cal M}_2^{p}(s, \tau) & = & \left(\frac{4}{9} u_v(s)  + \frac{1}{9} d_v(s) \right)   \times   \\
&&\hspace{-0.5 cm} \left( 1 + \frac{\tau}{4 \pi} C_{2NS}^{(1)}(s)+\left(\frac{\tau}{4 \pi}\right)^2 C_{2NS}^{(2)}(s)\right) e^{\tau
\Phi_{\rm NS}(s)},\nonumber
\end{eqnarray}
where, $C_{2NS}^{(1)}$  and $ C_{2NS}^{(2)}$ are the Wilson coefficients at NLO and NNLO accuracies in the Laplace $s$-space, respectively. Their analytical expressions are presented in appendix $A$.

For the non-singlet sector of  deuteron structure function, ${\cal M}_2^{d}$ (with $d = (p + n)/2$), in Laplace $s$-space {{} the relevant combination of parton densities}  at NNLO expansion can be written as
\begin{eqnarray} \label{eq:Fd}
{\cal M}_2^d(s, \tau) & = & \frac{5}{18} \, (u_v(s) + d_v(s)) \times \,,  \\
&& \hspace{-0.5 cm}  \left(1 + \frac{\tau}{4 \pi} C_{2NS}^{(1)}(s)+\left(\frac{\tau}{4 \pi}\right)^2 C_{2NS}^{(2)}(s)\right) e^{\tau
\Phi_{\rm NS}(s)}\;.\nonumber
\end{eqnarray}

It is now possible to consider the difference of proton  and deuteron  structure functions which is important to analysis the data in the region  $x \leq 0.3$. It  reads
\begin{eqnarray} \label{eq:FNS}
&{\cal M}_2^{\rm NS}(s, \tau)& \equiv 2 ( {\cal M}_2^{p} - {\cal M}_2^d)(s, \tau)  \\
&& =  \left( \frac{1}{3}
\,
(u_v - d_v)(s) + \frac{2} {3} \, (\bar u - \bar d)(s) \right) \times ~ \nonumber \\
&& \left(1 + \frac{\tau}{4 \pi} C_{2NS}^{(1)}(s)+\left(\frac{\tau}{4 \pi}\right)^2 C_{2NS}^{(2)}(s) \right)e^{\tau
\Phi_{\rm NS} (s)}\;.\nonumber
\end{eqnarray}

For $x<0.3$, the effect of sea quark densities {{}, $\bar u - \bar d$ in above equation,} can not be
neglected. In our calculations,  we take from {\tt JR14}~\cite{Jimenez-Delgado:2014twa} this distribution at the initial scale  Q$_0^2$ = 2 GeV$^2$ as,
\begin{equation}
x(\bar{d} - \bar{u})(x, Q_0^2) = 37.0 x^{2.2}(1 - x)^{19.2} (1 + 2.1 \sqrt{x})\;.
\end{equation}

For practical purposes, the combinations of $d - \bar d$ and $u - \bar u$  are also considered  as  the proton
valence densities. These are denoted as $d_v$  and $u_v$,  respectively.
The following valence distributions are employed in our analysis  at the input scale Q$_0^2$ = 2 GeV$^2$ \cite{Jimenez-Delgado:2014twa},
\begin{equation}\label{eq:xuvQ0}
	xu_v = {\cal N}_u x^{\alpha_{u_v}}(1-x)^{\beta_{u_v}}( 1 + {\gamma_{u_v}} x^{0.5} + {\eta_{u_v}}x )\,,
\end{equation}
\begin{equation} \label{eq:xdvQ0}
	xd_v = {\cal N}_d x^{\alpha_{d_v}}(1-x)^{\beta_{d_v}}( 1 + {\gamma_{d_v}} x^{0.5} + {\eta_{d_v}}x )\;,
\end{equation}
where,  ${\cal N}_u$ and ${\cal N}_d$ are the normalization factors which are given by
\begin{eqnarray}\label{eq:noal}
	{\cal N}_u &=& 2/ \left(B[\alpha_{u_v}, \beta_{u_v} + 1] + \eta_{u_v}B[\alpha_{u_v} +1, \beta_{u_v}+1]\right.
\nonumber \\ & + & \left. \gamma_{u_v} B[\alpha_{u_v} + 0.5, \beta_{u_v} + 1]\right) \,,   \nonumber  \\
	{\cal N}_d &=& 1/ \left(B[\alpha_{d_v}, \beta_{d_v} + 1] + \eta_{d_v}B[\alpha_{d_v} +1, \beta_{d_v}+1]\right.
\nonumber \\ & + & \left. \gamma_{d_v} B[\alpha_{d_v} + 0.5, \beta_{d_v} + 1]\right)   ,
\end{eqnarray}
where, the B-function stands for the  Euler beta function.

Using a variable change as  $x = e^{-\nu}$ and considering the Laplace transformation via the following relations
\begin{equation}
	u_v(s) = {\cal L}[e^{-\nu} u_v(e^{-\nu}); s] \,,
\end{equation}
\begin{equation}
	d_v(s) = {\cal L}[e^{-\nu} d_v(e^{-\nu}); s] \,,
\end{equation}
the valence distributions  in Eqs.~(\ref{eq:xuvQ0}) and (\ref{eq:xdvQ0}) can be presented in the Laplace $s$-space as
\begin{eqnarray}
	&& u_v(s)=2 (B[\alpha_{u_v}+s, \beta_{u_v}+1]+\eta_{u_v} B[\alpha_{u_v}+s+1, \beta_{u_v}+1]   \nonumber  \\
	&& + \gamma_{u_v} B[\alpha_{u_v}+s+0.5, \beta_{u_v}+1])/(B[\alpha_{u_v}, \beta_{u_v}+1]+    \nonumber   \\
	&& \eta_{u_v} B[\alpha_{u_v}+1, \beta_{u_v}+1] + \gamma_{u_v} B[\alpha_{u_v}+0.5, \beta_{u_v}+1])\;,
\end{eqnarray}
and
\begin{eqnarray}
	&& d_v(s)= (B[\alpha_{d_v}+s, \beta_{d_v}+1]+\eta_{d_v} B[\alpha_{d_v}+s+1, \beta_{d_v}+1]   \nonumber  \\
	&& + \gamma_{d_v} B[\alpha_{d_v}+s+0.5, \beta_{d_v}+1])/(B[\alpha_{d_v}, \beta_{d_v}+1]+    \nonumber   \\
	&& \eta_{d_v} B[\alpha_{d_v}+1, \beta_{d_v}+1] + \gamma_{d_v} B[\alpha_{d_v}+0.5, \beta_{d_v}+1])\;.
\end{eqnarray}

To extract the unknown parameters in the valance distribution, one needs to do a global fit based on the Jacobi polynomial approach.
Details of this approach can be found in Refs.~\cite{Khorramian:2008yh,Khorramian:2009xz}.
Here, we review this method briefly. Using this approach, the structure functions can be reconstructed  as
\begin{eqnarray}\label{eq:F_2Jacobi}
&&F_2^{p,d,NS}(x, Q^2)  =  x^{\beta}(1 - x)^{\alpha} \, \sum_{n=0}^{N_{max}} \, \Theta_n^{\alpha, \beta}(x) \nonumber
\\
&&\times  \sum_{j=0}^n \, c_j^{(n)}{(\alpha, \beta)} \,{{\cal L}} [F_2^{p,d,NS}, j + 1] \;.
\end{eqnarray}

In the above equation, the Jacobi polynomials $\Theta_n^{\alpha, \beta}(x)$ have the following expansion
\begin{equation}\label{eq:Jacobi-polynomials}
\Theta_n^{\alpha, \beta}(x) = \sum_{j = 0}^{n} \, c_j^{(n)}(\alpha, \beta) \, x^j \,,
\end{equation}
where, $c_j^{(n)}(\alpha, \beta)$ are the coefficients which can be written in terms of the Euler Gamma function. The  parameters $\alpha$ and
$\beta$ are fixed and taken to be  3 and 0.7, respectively \cite{Khorramian:2008yh,Khorramian:2009xz}.
As a final point, by considering the weight function $x^{\beta} (1-x)^{\alpha}$ the orthogonality condition  for the  Jacobi polynomials is given by
\begin{equation}\label{eq:orthogonality-relation}
\int_0^1 dx \, x^{\beta} (1 - x)^{\alpha} \, \Theta_{k}^{\alpha, \beta}(x) \, \Theta_{l}^{\alpha, \beta}(x) =
\delta_{k,l} \,.
\end{equation}

The moment of structure function in Laplace $s$-space, given by last term in Eq.~(\ref{eq:F_2Jacobi}), can be written in terms of Eq.~(\ref{eq:Fp}). The global fit over all available experimental data for the structure functions will provide us the unknown parameters of  valance densities. Technical details of the global fit are investigated in the following section.

\begin{figure}
	\includegraphics[clip,width=0.35\textwidth]{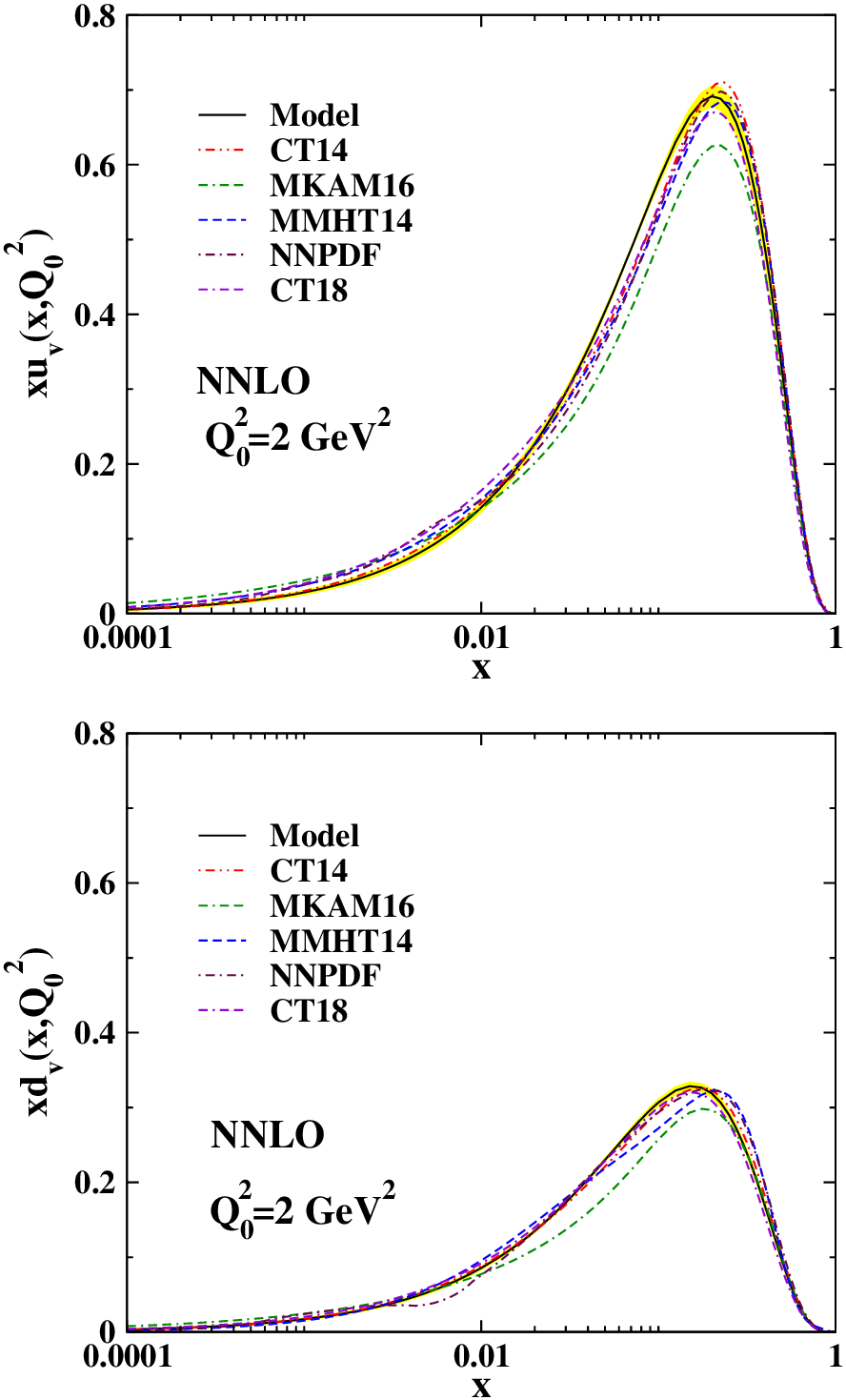}
	\begin{center}
		\caption{(Color online) The up and down valence parton distributions $x u_v$ and $x d_v$ in NNLO accuracy  with $\Delta \chi^2=1$ uncertainty bands  at the initial scale $Q_0^2 = 2$~$GeV^2$. The dashed-dashed-dotted curve stands for the {\tt MKAM16} model~\cite{MoosaviNejad:2016ebo}, the dashed-dotted one shows  the {\tt NNPDF} model~\cite{Ball:2012cx}, the long-dashed curve refers to  the {\tt MMHT14} model~\cite{Harland-Lang:2014zoa}, the dashed-dotted-dotted curve shows the {\tt CT14} model~\cite{Dulat:2015mca} and the dashed-dashed-dotted curve represents the {\tt CT18} model~\cite{Hou:2019efy}.}\label{fig:xuvandxdvnnlo}
	\end{center}
\end{figure}

\begin{figure}[htb]
	\begin{center}
		\includegraphics[clip,width=0.38\textwidth]{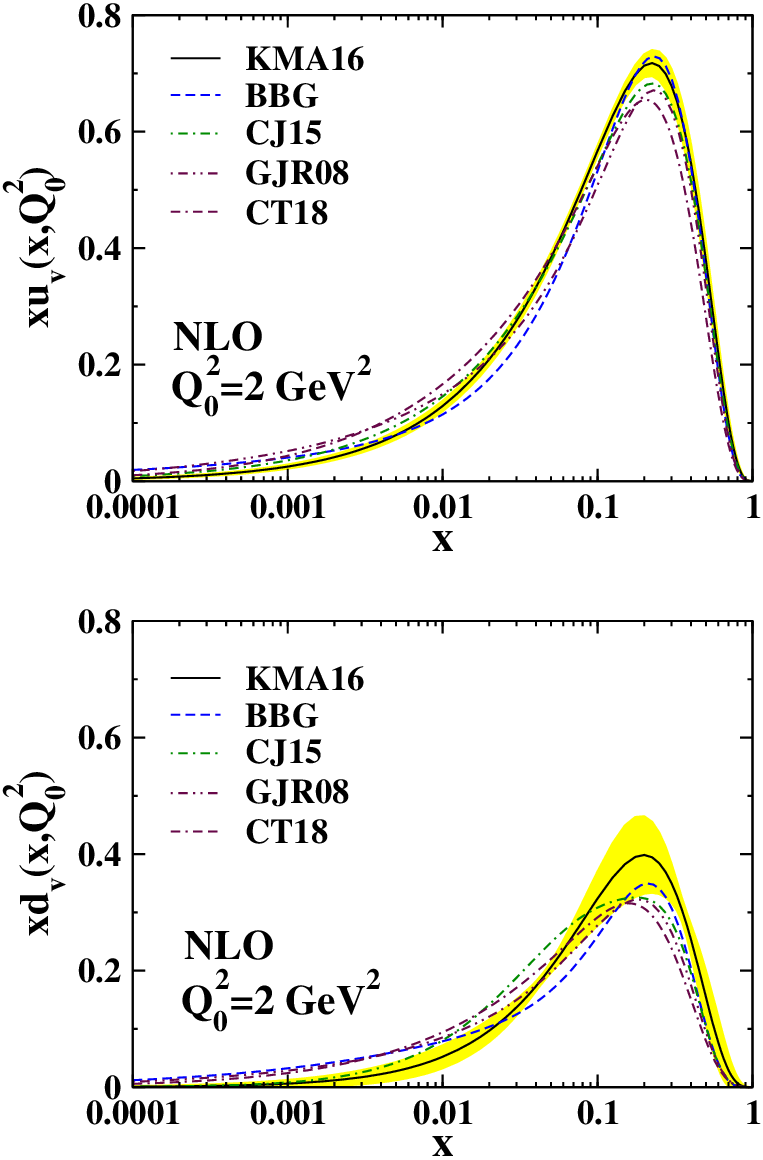}
		\caption{(Color online) The parton densities $xu_v$ and $xd_v$ in NLO accuracy  at the input scale Q$^2_0= 2 \, {\rm GeV}^2$. The solid line is the {\tt KMA16} PDF~\cite{Khanpour:2016uxh}, dashed 	line is the {\tt BBG} PDF~\cite{Blumlein:2006be}, the dashed-dotted line is the {\tt CJ15} PDF~\cite{Accardi:2016qay}, dashed-double-dotted line is the {\tt GJR08} PDF~\cite{Gluck:2008gs} and dashed-dashed-dotted line shows the result from {\tt CT18} model~\cite{Hou:2019efy}.}\label{fig:partonvalnceQ0}
	\end{center}
\end{figure}

Numerical results for the free parameters in the  $u_v$  and $d_v$ densities  at  NNLO and NLO accuracies   are listed in Tables.~\ref{tab:fitresultsNNLO} and \ref{tab:fitresultsNLO}, respectively. They are resulted from non-singlet QCD fit at $Q^2_0=2\;GeV^2$, as would be described in Sec.~\ref{sec:dataset}.  Although, the uncertainty bands of PDFs are related to the experimental uncertainties but they are also affected by increasing the theoretical precision.  As is seen from Figs.~\ref{fig:xuvandxdvnnlo} and \ref{fig:partonvalnceQ0} , the error bands for NNLO results are smaller than the NLO one which indicates an enhancement in accuracy of analysis.
\begin{table}
	\begin{tabular}{c | c | c }
		\hline
		\multicolumn{3}{c}{ NNLO non-singlet QCD fit }   \\ \hline
		\hline $u_v$      & $\alpha_u$         &  0.70639 $\pm$ 0.0205  \\
		& $\beta_u$                    &  3.5318  $\pm$ 0.0183   \\
		& $\gamma_u  $                  &  1.000              \\
		& $\eta_u$                    &  1.1400                 \\
		\hline $d_v$     & $\alpha_d$          & 0.71113 $\pm$ 0.0173  \\
		& $\beta_d$                      &  4.2114  $\pm$ 0.0955    \\
		& $\gamma_d  $                    & 1.5999                 \\
		& $\eta_d$                      &  4.2899                \\
		\hline \multicolumn{2}{c|}{ $\alpha_{\rm s}^{\rm N_f=4}(Q_0^2)$}
		& 0.3632 $\pm$ 0.0149 \\
		\hline \hline \multicolumn{2}{c|}{ $\chi^2 / {\rm n.d.f}$ } & 510.113/563 = 0.906  \\
		\hline \hline
	\end{tabular}
	\caption{\label{tab:fitresultsNNLO}{\sf Parameter values of NNLO non-singlet QCD fit at Q$_0^2 = 2 \, {\rm GeV}^2$.
	}}\label{nnlo}
\end{table}
\begin{table}
	\begin{tabular}{c | c | c }
		\hline
		\multicolumn{3}{c}{ NLO non-singlet QCD fit }   \\ \hline
		\hline $u_v$      & $\alpha_u$         &  0.7108 $\pm$ 0.1295  \\
		& $\beta_u$                    &  3.3595  $\pm$ 0.027   \\
		& $\gamma_u  $                  &  0.2979               \\
		& $\eta_u$                    &  1.3440                 \\
		\hline $d_v$     & $\alpha_d$          &  0.9467 $\pm$ 0.0261  \\
		& $\beta_d$                      &  2.8468  $\pm$ 0.3130    \\
		& $\gamma_d  $                    &  1.1004                 \\
		& $\eta_d$                      &  -1.1330                \\
		\hline \multicolumn{2}{c|}{ $\alpha_{\rm s}^{\rm N_f=4}(Q_0^2)$}
		& 0.3521 $\pm$ 0.0139 \\
		\hline \hline \multicolumn{2}{c|}{ $\chi^2 / {\rm n.d.f}$ } & 521.303/563 = 0.92  \\
		\hline \hline
	\end{tabular}
	\caption{\label{tab:fitresultsNLO}{\sf Parameter values of NLO non-singlet QCD fit at Q$_0^2 = 2 \, {\rm GeV}^2$}~\cite{Khanpour:2016uxh}.}\label{nlo}
\end{table}

\section{ Global analyses of valence-quarks densities }
\label{sec:dataset}
\subsection{ Different features of data sets }

The valence PDFs can be determined by fitting to a global data set  over 572 data points including a variety range of scattering processes at high energies.
In Table.~\ref{tab:data}, the data sets used in our analysis are listed.  The DIS data from {\tt 	BCDMS}~\cite{Benvenuti:1989fm,Benvenuti:1989rh,Benvenuti:1989gs} and  {\tt SLAC}~\cite{Whitlow:1991uw} as well as  {\tt 	NMC}~\cite{Arneodo:1996qe,Arneodo:1995cq} experiments make the required sets. The flavor separation of PDFs at large $x$ is facilitated by these data sets. In our analysis,  the DIS data from {\tt H1}~\cite{H1} and {\tt ZEUS}~\cite{ZEUS} Collaborations are also employed. Additionally, the combined measurements of H1 and ZEUS Collaborations at HERA for the inclusive $e^\pm p$ scattering cross sections are applied as the new data sets \cite{Aaron:2009aa}. For the  valence quarks in the region $x \geq 0.3$, we use the  data set given for $F_2^p (x, Q^2)$ and $F_2^d (x, Q^2)$ while for the region $x < 0.3$ the data are related to $F_2^{\rm NS}(x, Q^2) = 2 (F_2^p(x, Q^2) - F_2^d(x, Q^2))$.

In order to eliminate the higher twist (HT) effects, it is needed  to take  different cuts on data sets  before analysis. The DIS data cuts are considered for  Q$^2>4$~GeV$^2$ and on the hadronic mass  $W^2>12.5$~GeV$^2$ which perform the required converge in the concerned kinematic region. For the  {\tt BCDMS}   and  {\tt NMC}   data sets, additional cuts as $y>0.35$ and $Q^2>8$~GeV$^2$
should be applied, respectively. Considering the required cuts, we  listed in
Table.~\ref{tab:data}  both the required DIS data and the number of data points for each experiment in the fitting process.  In the 5$^{\rm th}$ column of Table.~\ref{tab:data}, taking the additional cuts the number of reduced data points are  presented. Consequently, the number of data points are reduced from 467 to 248
for $F_2^{p} (x, Q^2)$ so that for $F_2^d (x, Q^2)$ this reduction is from 232 to 159. At last,  for $F_2^{\rm NS} (x, Q^2)$ the number of data points are reduced from 208 to 165.
\renewcommand{\arraystretch}{0.8}
\begin{table*}
	\caption{  \label{tab:data}{  \sf Data sets used in our QCD analysis for (a) $F_2^p (x, Q^2)$, (b) $F_2^d (x, Q^2)$, and (c) $F_2^{\rm NS} (x, Q^2)$. The name of different data sets as well as the range of $x$ and Q$^2$ are given in three first
			columns. The normalization shifts are also listed in the last column. }}
		\begin{tabular}{||l|c|c|c|c||c||}
		\hline
		{\tt Experiment}  & $x$ & Q$^2 ({\rm GeV}^2)$ & $F_2^p$ & $F_2^p~{\rm cuts}$ &  ${\cal N}$  \\  \hline
		\hline {\tt BCDMS (100)}   & 0.35--0.75 &  11.75--75.00  & 51 & 29 	& 0.996805 \\
		{\tt BCDMS (120)}   & 0.35--0.75 &  13.25--75.00  &  59 &  32 &  0.996805      \\
		{\tt BCDMS (200)}   & 0.35--0.75 &  32.50--137.50  &  50 &  28 & 	0.997833   \\
		{\tt BCDMS (280)}   & 0.35--0.75 &  43.00--230.00  &  49 &  26 & 	1.002131    \\
		{\tt NMC (comb)}    & 0.35--0.50 &  7.00--65.00  &  15 &  14 & 	1.000158       \\
		{\tt SLAC (comb)}   & 0.30--0.62 &  7.30--21.39  &  57 &  57 &  1.000714       \\
		{\tt H1 (hQ2)}      & 0.40--0.65 &  200--30000  &  26 &  26 & 	1.001000       \\
		{\tt ZEUS (hQ2)}    & 0.40--0.65 &  650--30000  &  15 &  15 &   0.999929             \\
		{\tt H1 (comb)}    &  0.40--0.65 &  90--30000  &  145 &  21 & 	0.999947        \\
		\hline
		{\bf proton}      &              &                    & 467 & 248 &          \\
		\hline
	\end{tabular}
	\\
	\vspace{0.1cm} \text{(a)  $F_2^p (x, Q^2)$ data points~\cite{Benvenuti:1989fm,Benvenuti:1989rh,Benvenuti:1989gs,Whitlow:1991uw,Arneodo:1996qe,Arneodo:1995cq,H1,ZEUS,Aaron:2009aa}.} \\
	\vspace{0.3cm}
	\begin{tabular}{||l|c | c |c |c ||c||}
		\hline
		{\tt Experiment}  & $x$ & Q$^2 ({\rm GeV}^2)$ & $F_2^d$ & $F_2^d~{\rm cuts}$ & ${\cal N}$  \\   \hline
		\hline
		{\tt BCDMS (120)}   & 0.35--0.75 & 13.25--99.00   &  59 &  32 &    1.007303 \\
		{\tt BCDMS (200)}   & 0.35--0.75 & 32.50--137.50  &  50 &  28 &    1.001829 \\
		{\tt BCDMS (280)}   & 0.35--0.75 & 43.00--230.00  &  49 &  26 &    1.001742 \\
		{\tt NMC (comb)}    & 0.35--0.50 & 7.00--65.00    &  15 &  14 &    0.998725 \\
		{\tt SLAC (comb)}   & 0.30--0.62 & 10.00--21.40   &  59 &  59 &    0.997742 \\
		\hline
		{\bf deuteron}    &              &                  & 232 & 159 &  \\
		\hline
	\end{tabular}
	\\
	\vspace{0.1cm} \text{(b)  $F_2^d (x, Q^2)$ data points~\cite{Benvenuti:1989fm,Whitlow:1991uw,Arneodo:1996qe,Arneodo:1995cq}.} \\
	\vspace{0.3cm}
	\begin{tabular}{||l|    c   | c |c |c ||c||}
		\hline  {\tt Experiment}  & $x$ & Q$^2 ({\rm GeV}^2)$ & $F_2^{\rm NS}$ & $F_2^{\rm NS}~{\rm cuts}$
		& ${\cal N}$  \\
		\hline    \hline
		{\tt BCDMS (120)}  & 0.070--0.275 &  8.75--43.00    &  36 &  30 &     0.998751  \\
		{\tt BCDMS (200)}  & 0.070--0.275 &  17.00--75.00   &  29 &  28 &     0.998758  \\
		{\tt BCDMS (280)}  & 0.100--0.275 &  32.50--115.50  &  27 &  26 &     0.999529  \\
		{\tt NMC (comb)}   & 0.013--0.275 &  4.50--65.00    &  88 &  53 &     1.000135  \\
		{\tt SLAC (comb)}  & 0.153--0.293 &  4.18--5.50     &  28 &  28 &     1.000923  \\
		\hline
		{\bf non-singlet} &               &                 & 208 & 165 &    \\
		\hline
	\end{tabular}
	\\
	\vspace{0.1cm} \text{(c) $F_2^{\rm NS} (x, Q^2)$ data points~\cite{Benvenuti:1989fm,Whitlow:1991uw,Arneodo:1996qe,Arneodo:1995cq}. \vspace{0.5cm}}
\end{table*}

\subsection{$\chi^2$-Minimization approach}
The best values of  fit parameters at NNLO accuracy  are  extracted by minimizing the  $\chi^2$ with respect to four unknown parameters in valence-quark distributions (\ref{eq:xuvQ0}) and (\ref{eq:xdvQ0}),  along  with $\Lambda_{\rm{\overline {MS}}}^{(4)}$ as the QCD cutoff parameter.
As is shown in Tables.~\ref{nnlo} and \ref{nlo}, we fix four parameters in the fit from the beginning, i.e. $\gamma_u, \eta_u, \gamma_d$ and $\eta_d$, so there are totally five free parameters remaining to be extracted from the fit process.\\
The global goodness-of-fit can be done using the following definition for the effective $\chi^2$ \cite{cor}
	\begin{eqnarray}
		\chi _{\mathrm{global}}^{2} =
		\sum_{n} w_{n} \chi _{n}^{2}\;,
	\end{eqnarray}
where, $ n$ labels  the different experiments and
	\begin{eqnarray}
		\chi _{n}^{2} =\left(\frac{1-{\cal N}_{n}}{\Delta{\cal
				N}_{n}}\right)^{2} +\sum_{i}\left( \frac{{\cal
				N}_{n}F_{2,i}^{data}-F_{2,i}^{theor}}{{\cal N}_{n}\Delta
			F_{2,i}^{data}} \right)^{2}. \label{eq:Chi2n}
	\end{eqnarray}
In the above relation,  $F_{2,i}^{data}$, $\Delta F_{2,i}^{data}$  and $F_{2,i}^{theor}$ stand for  the data value, the data uncertainty including a combination of statistical and systematic errors, and the theoretical value for 	the $i^{\mathrm{th}}$ data point in the $n^{\mathrm{th}}$ experiment, respectively. Moreover, ${\Delta{\cal N}_{n}}$ in the first parenthesis  is the experimental normalization uncertainty  and for the data of $n^{\mathrm{th}}$ experiment  an 	overall normalization factor is given by  ${\cal N}_{n}$.  	This factor is extracted through the fit procedure and when it is determined through the first fit we fix its value in next stages. The possible weighting factor denotes by  $w_{n}$  which is usually taken to be 1. 	In our analysis, a relative normalization shift   between the different data sets, i.e., ${\cal N}_{n}$,  is used for which  the normalization 	uncertainties ${\Delta{\cal N}_{n}}$ are given by the experiments.
		
In our analysis, the total number of data points are $n^{data}=572$ and there are also five unknown parameters in the fit. The best parametrization of valence-quark densities are obtained using the CERN program library {\tt MINUIT}~\cite{CERN-Minuit}. 
The results  are listed in Tables.~\ref{tab:fitresultsNNLO} and \ref{tab:fitresultsNLO}.
\subsection{ Determining the input uncertainties}
%
In this subsection, we describe our method to determine the uncertainties of valence-quark PDFs as well as the error propagation from experimental data points. {{} In practice, the uncertainties in global PDF analysis can be achieved by procedures which are well-defined to propagate through the experimental uncertainties on the
fitted data points to the PDF uncertainties.}
For this purpose, we apply the Hessian method (or error matrix approach)    \cite{Pumplin:2001ct} which is
based linearly on error propagation  and  the suitable production of  PDF eigenvector. The Hessian method was firstly used in  analysis by MRST04~\cite{Martin:2003sk}, MSTW08~\cite{Martin:2009iq}, MRST03~\cite{Martin:2002aw}, CETQ group \cite{Dulat:2015mca} and its up-to-dated version by CT18 \cite{Hou:2019efy}, and also in our previous works
\cite{AtashbarTehrani:2012xh,Khanpour:2016pph,Monfared:2011xf,Khanpour:2012tk}. By  running the
CERN program library MINUIT~\cite{CERN-Minuit}, an error analysis can be done  using the Hessian matrix. A
simple and efficient method to calculate the uncertainties of parton densities can be obtained by applying this method which is essentially related to
diagonalize the covariance matrix. In the  Hessian approach, the main assumption is to perform  a quadratic expansion of $\chi^2_{\text{ global}}$ with respect to the  fitted  parameters $a_i$ about its global
minimum. It reads
\begin{equation} \label{chi2-2}
\Delta \chi^2 \equiv  \chi^2_{\text{global}}  - \chi^2_{\text min} = \sum_{\text{ i, j=1}}^p (a_i -
a_i^0) \, H_{\rm ij} \,   (a_j - a_j^0) \;.
\end{equation}
In this equation,  the elements of the Hessian matrix are denoted by $H_{\rm ij}$  and
the  number of parameters in the global fit is presented by $p$.

\subsection{ Error calculations }
\begin{figure}
	\begin{center}
		\vspace{1cm}
		\resizebox{0.49\textwidth}{!}{\includegraphics{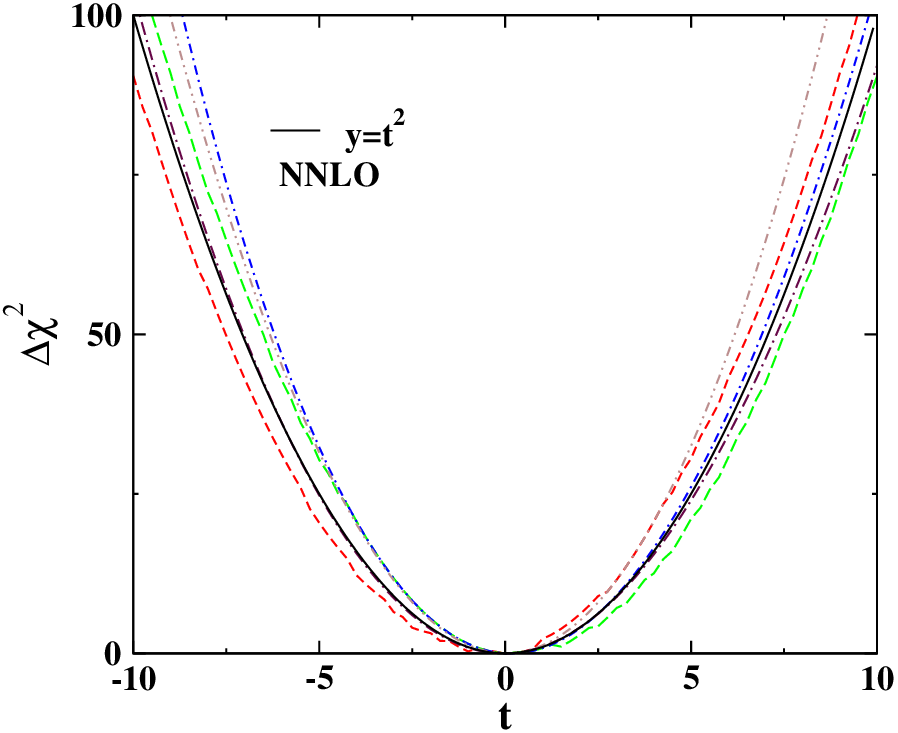}}
		\caption{(Color online)  $\Delta \chi^2$ as a function of $t$ at  NNLO accuracy~\cite{Khanpour:2016pph,AtashbarTehrani:2012xh,Martin:2002aw,Pumplin:2001ct,Martin:2003sk,Martin:2009iq}.{}{ The indicated
				curves correspond to all five eigenvectors.}}\label{fig:deltachinnlo}
	\end{center}
\end{figure}

%
The method presented in
Refs.~\cite{Martin:2002aw,Martin:2003sk,Pumplin:2001ct,Martin:2009iq,AtashbarTehrani:2012xh,Khanpour:2016pph} can be followed to determine the error of calculations.
The eigenvectors and eigenvalues of the covariance (or Hessian) matrix are the basic mathematical tools. For practical purposes we need, at first,  to have a set of appropriate fit parameters considered in the valence quark  densities. {{} They are usually labeled by $a_i(s_{\text {min}})$ and required}  to minimize the $\chi^2_{\text{ global}}$. Considering the parton sets as  $s_k^\pm$, it is possible to have
an
expansion in terms of the  eigenvectors and eigenvalues for the variation of parameters around the global minimum, i.e.
\begin{equation}
a_i (s_k^\pm)  = a_i (s_{\text{min}})  \pm  t \sqrt{\lambda_k} v_{\text {ik}} \;. \label{aaa}
\end{equation}
Here,  $a_i (s_k^\pm)$ and $a_i (s_{\text{min}})$ {{} are the same as  the  abbreviated  $a_i$ and  $a_i^0$ symbols} in Eq.~(\ref{chi2-2}).
In the above equation, the k$^{\text {th}}$-eigenvalue and the i$^{\text {th}}$-component of the
orthonormal eigenvectors of Hessian matrix are denoted by $\lambda_k$ and $v_{\text {ik}}$, respectively.
{{} By adjusting the $t$ variable proportionally, it is possible to set  $t = T$ in the quadratic approximation and therefore to get $T^2 = \Delta \chi^2$.} The quadratic
approximation of Eq.~(\ref{chi2-2}) can be tested by considering  the dependence of
$\Delta \chi^2$ on the  eigenvector directions for some selected samples. {{} In Fig.~\ref{fig:deltachinnlo},  $\Delta \chi^2$ is plotted as a {{} function of $t$ for all eigenvectors}  at the  NNLO accuracy.}

\begin{table*}[htb]
	\renewcommand{\arraystretch}{1.30}
	\centering
	{\footnotesize
		\begin{tabular}{||c||c|c|c|c|c||}
			\hline \hline
			& $\alpha_{u_v}$ & $\beta_{u_v}$ &$\alpha_{d_v}$ & $\beta_{d_v}$ & $\Lambda_{\overline{\rm MS}}^{(4)}$
			\\
			
			\hline \hline
			$\alpha_{u_v}$       & 1 &  &  &  &   \\
			\hline
			$\beta_{u_v}$       & 0.0957708  & 1  &  &  & \\
			\hline
			$\alpha_{d_v}$       &-0.0500717 & -0.294017 & 1   & &  \\
			\hline
			$\beta_{d_v}$       &-0.0315322  &-0.23711 &   0.910278&
			1     & \\
			\hline
			$\Lambda_{\overline{\rm MS}}^{(2)}$& 0.0445004 &-0.585629 & -0.0247452 & -0.127968   & 1   \\
			\hline\hline
		\end{tabular}
	}
	\caption[]{ The correlation matrix elements for the 4 + 1 free parameters at  NNLO fit.}
	\label{covmat-matrixNNLO}
\end{table*}

In our NNLO global analysis, the  covariance matrix elements for five free parameters are given in Table.~\ref{covmat-matrixNNLO}. The estimation for uncertainties of parton densities which is generally denoted by $F$  can be determined, using the following relation introduced in \cite{Martin:2002aw}:
\begin{equation}
(\Delta F)^2 = \Delta\chi^2
\sum_{i=1}^{p}\sum_{j=1}^p \frac{\partial F}{\partial a_i}
C_{ij}(a)\frac{\partial F}{\partial a_i},
\label{eq:covariance}
\end{equation}
where,  $C_{ij}(a) =  (H^{-1})_{ij}$ is the covariance matrix and $\Delta \chi^2$  is the allowed variation in
$\chi^2$, as was specified previously. By  suitable choice of $\Delta \chi^2=1$, which corresponds to one sigma confidence level, and by considering  the Hessian (or equivalently  covariance)
matrix it is now possible to calculate the errors on parton densities.

Now, the valence quark densities can be calculated  at higher-Q$^2$
(i.e. $Q^2>Q_0^2$)  using the DGLAP evolution equations in Laplace $s$-space.
Using the  analytical solution, based on  the Laplace transform technique as well as Jacobi polynomials approach, the  global QCD analysis can be done. In Table.~\ref{nnlo}, we listed  the numerical  values of PDFs parameters  which are related to the QCD fit for the  NNLO non-singlet sector at the input scale Q$_0^2 = 2$~GeV$^2$.
After the first minimization, some parameters take values with very small error so that one can ignore their errors and  fix them to their central values through the fit procedure. By fixing some parameters, the fit is repeated which {{} could be done  easier and faster. Hence some parameters are presented  in  Tables.~\ref{nnlo} and \ref{nlo}  without any error}. We can see from Table.~\ref{nnlo} that the central values of PDF parameters are rather stable.  From the fit, one can find the numerical value  for
$\alpha_{\rm s}^{\rm N_f=4}(Q_0^2)$  as presented in Tables.~\ref{nnlo} and \ref{nlo}. This implies the numerical value for QCD cutoff quantity. Using this quantity and  considering the threshold matching condition for running coupling constant \cite{match}, we get $\alpha_s(M_Z^2) = 0.1147 \pm 0.0009$  at the mass scale of $Z$-boson.
In Fig.~\ref{fig:xuvandxdvnnlo}, we depicted the $xu_v$- and $xd_v$-distributions at the energy scale $Q_0^2=2$~GeV$^2$. This also  contains the  PDF uncertainties with required confidence level corresponding to  $\Delta \chi^2=1$  in Eq.~(\ref{chi2-2}). In this figure, we have also added  the results from other models such {\tt MKAM16}~\cite{MoosaviNejad:2016ebo}, {\tt NNPDF}~\cite{Ball:2012cx}, {\tt MMHT14}~\cite{Harland-Lang:2014zoa}, {\tt CT14}~\cite{Dulat:2015mca} and up-to-dated results from {\tt CT18} PDFs~\cite{Hou:2019efy} to have a more qualitative comparison. To confirm the considerable progress of PDFs at NNLO accuracy, the results of NLO PDFs for up and down valence densities have  been also depicted in  Fig.~\ref{fig:partonvalnceQ0}  at the scale $Q_0^2=2$~GeV$^2$. The  numerical values of fitted parameters at NLO accuracy are listed in Table.~\ref{nlo}. As is seen from Tables.~\ref{nnlo} and \ref{nlo}, the value of $\chi^2 / {\rm n.d.f}$ is reduced from 0.920 to 0.906 {{} by increasing the accuracy of corrections from NLO to NNLO.}
Moreover, in Fig.~\ref{fig:partonvalnceQ0} the small uncertainty raised in our NNLO analysis confirms the validity of NNLO analysis in comparison with the NLO one.  
Of course,  the higher order correction   is not the only reason for decreasing the errors and additionally the type of fitting  procedure   would be also effective on getting the small error bands.
This point will be described in the following section.
\section{ Higher twist and target mass corrections }
\label{Sec:TMC}
\begin{figure*}[htb]
	\begin{center}
		\includegraphics[clip,width=0.5\textwidth]{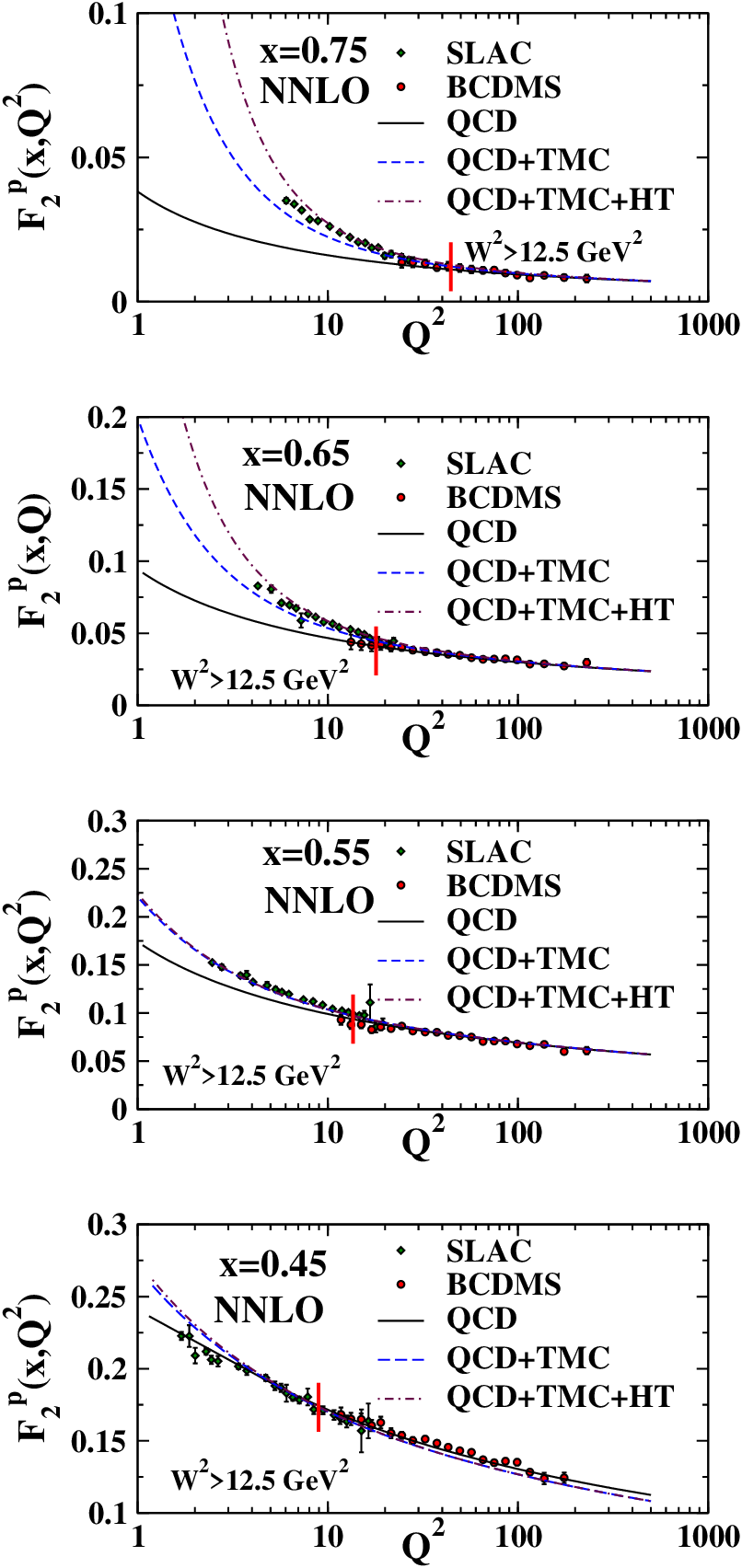}
		\caption{ (Color online) Comparison of data for proton structure function ($F^{p}_2$) from
			{\tt BCDMS}~\cite{Benvenuti:1989fm,Benvenuti:1989rh,Benvenuti:1989gs} and {\tt SLAC}~\cite{Whitlow:1991uw}, with our
			theoretical predictions as a function of Q$^2$ for fixed values of $x$. {{} This plot  includes, the pure QCD fit at NNLO (solid line), the TMCs contributions (dashed line), and the HT effect (dashed-dotted line).}   \label{fig:f2p}}
	\end{center}
\end{figure*}
\begin{figure*}[htb]
	\begin{center}
		\includegraphics[clip,width=0.5\textwidth]{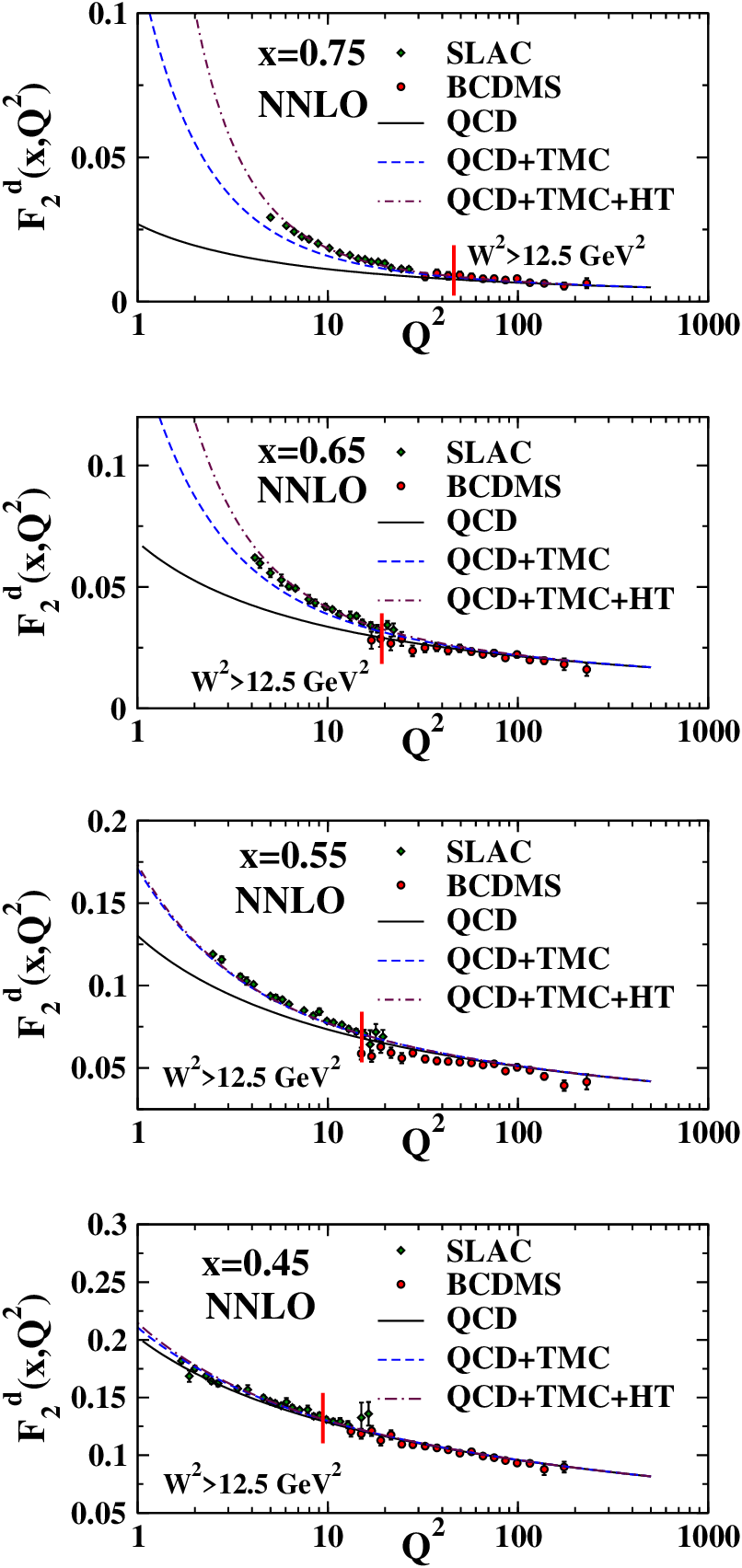}
		\caption{ (Color online) As in Fig.~\ref{fig:f2p}, but for  deuteron structure function ($F^d_2$). {{} In this figure as well as  Fig.~\ref{fig:f2p}, to extract the PDFs parameters through the fit we just used  the data points after the vertical red lines. These data points belong to the region $W^2\geq 12.5\; GeV^2$.}
			\label{fig:f2d}}
			\end{center}
\end{figure*}
To determine the moment of structure function, it is assumed that in the limit of
$Q^2\rightarrow\infty$ the  mass of target hadron approaches to zero. This assumption  would be violated at the  intermediate and low values of $Q^2$. In these regions, the
moment is significantly  affected by the  power correction of order ${\cal O}(m_N^2/Q^2)$ where $m_N$ denotes the nucleon mass \cite{Kataev:1996vu,Kataev:1997nc}.
This correction is known as the target
mass correction (TMC).  Taking into account the TMC effect leads to change the moment of non-singlet structure function as it follows
~\cite{Khorramian:2008yh,Khorramian:2009xz,Khanpour:2016uxh,Georgi:1976ve,Gluck:2006yz,Steffens:2012jx},
\begin{eqnarray}\label{equ:TMC}
&& {\cal M}_{2,{\rm TMC}}^{k}(s, Q^2)     \equiv
{\cal L} [{\cal M}_{2, {\rm TMC}}^k(e^{-v}, Q^2; s)]\,   \nonumber  \\
&& = {\cal M}_2^k(s, Q^2)+\frac{s (s-1)}{s+2}\left(\frac{m_N^2}{Q^2}\right) \, {\cal M}_2^{
	k}(s+2, Q^2)   \nonumber \\
&&+\frac{\Gamma(s+3)^2}{2\Gamma(s-1)\Gamma(s+5)}\left(\frac{m_N^2}{Q^2}\right)^2 \, {\cal M}_2^{
	k}(s+4, Q^2)   \nonumber  \\
&&+\frac{\Gamma(s+4)\Gamma(s+5)}{6\Gamma(s-1)\Gamma(s+7)}\left(\frac{m_N^2}{Q^2}\right)^3 \, {\cal M}_2^{
	k}(s+6, Q^2)   \nonumber  \\
&&+\frac{\Gamma(s+5)\Gamma(s+7)}{24\Gamma(s-1)\Gamma(s+9)}\left(\frac{m_N^2}{Q^2}\right)^4 \, {\cal M}_2^k(s+8, Q^2)
\nonumber  \\
&&+ {\cal{O}}\left( \frac{m_N^2}{Q^2}\right)^5 \;.
\end{eqnarray}

In the relevant region where  $x < 0.8$, the higher powers of $(m_{\rm N}^2/Q^2)^n$ (for $n \geqslant 2$) are negligible \cite{Georgi:1976ve}.
By substituting Eq.~\eqref{equ:TMC} into Eq.~\eqref{eq:F_2Jacobi}, one  obtains
\begin{eqnarray} \label{eg1JacobTMC}
F_2^{k, {\rm TMC}}(x,Q^2) & = & x^{\beta}(1 - x)^{\alpha} \sum_{n = 0}^{\rm N_{max}} \Theta_n ^{\alpha, \beta}(x)
   \\
& \times & \sum_{j = 0}^n \, c_{j}^{(n)}{(\alpha, \beta)} \, {\cal M}_{2, {\rm TMC}}^k (j+1, Q^2)\,, \nonumber
 \end{eqnarray}
where, ${\cal M}_{2, {\rm TMC}}^k(j+1, Q^2)$ is the moment of concerned  structure function in the Laplace $s$-space. This term contains the
TMC effect and is given
by Eq.~\eqref{equ:TMC} where the  corrections due to higher powers (for $n \geqslant 2$) are neglected.

The effect of higher twist (HT) correction is also significant at large values of $x$ and moderate $Q^2$~\cite{Abt:2016vjh,Jimenez-Delgado:2013boa,Leader:2006xc,Nath:2016phi,Wei:2016far} where the TMC effect is also considerable.
Consequently, inclusion of higher twists has its own importance to analysis the  parton densities.

The required kinematic cuts, containing the HT effect for $F_2^{p}(x, Q^2)$
and $F_2^d(x, Q^2)$, are  as $Q^2 \geq 4$~GeV$^2$ and $W^2 \geq 12.5$~GeV$^2$ where $W^2$ denotes the hadronic invariant mass.
The cuts can be extended to  the kinematic  region  $4 < W^2 < 12.5 {\rm
GeV}^2$.
Therefore, an extrapolation is needed to this region in our  QCD fit. It should be noted that, the $W^2$ is related to the $Q^2$-variable through $W^2=(1/x-1)Q^2+m^2_n$, where $m_n$ is the nucleon mass. From this relation, it can be seen that the
required kinematic cut is such that for $W^2 \geq 12.5$ GeV$^2$, the $Q^2$ is also increasing. Consequently, the TMC and HT effects can be
ignored in this region. Therefore, as is seen from Figs.~\ref{fig:f2p} and \ref{fig:f2d}, three plots labeled as QCD, QCD+TMC and QCD+TMC+HT show the same behavior.

Concerning the higher twist effect it should be noted that, the parametrization form of  higher twist contributions are practically  independent of the leading twist one. This parametrization is
typically given by polynomial functions of $x$. For the  DIS data where  the power corrections in the concerned region can not be neglected, the
corrections are defined by an ansatz which is completely motivated from  phenomenological points of view and it reads  \cite{Blumlein:2008kz,Blumlein:2006be,Martin:1998np,Martin:2003sk}
\begin{eqnarray}\label{equ:HTC}
F_2^{\rm HT}(x, Q^2) = {\cal O}_{\rm TMC}[F_2^{\rm TMC}(x, Q^2)]  \left( 1 + \frac{h(x, Q^2)}{Q^2[{\rm GeV}^2]}
\right)\,, \nonumber \\
\end{eqnarray}
where $F_2^{\rm TMC}(x,Q^2)$ is given by Eq.~\eqref{eg1JacobTMC}. The operation ${\cal O}_{\rm TMC} [\cdots]$ in the above equation refers to the target
mass correction while the twist-2 contribution is considered for the desired structure function. It should be noted that, the  $h(x, Q^2)$-coefficient is
determined at individual intervals of $x$ and Q$^2$ but finally the averaged result over Q$^2$ is taken into account. Therefore,  the $x$-dependence of
$h(x)$ as the higher twist contribution could be defined as \cite{Blumlein:2008kz}
\begin{equation}
h(x) = \alpha \left (\frac{x^\beta}{1 - x} - \gamma \right)\label{hx} \,.
\end{equation}

To achieve a  flexible higher twist contribution during the data analysis,  the above choice of
$h(x, Q^2)$ would be effective. As was mentioned, the cuts $Q^2 \geq 4 \, {\rm GeV}^2$
and $4 < W^2 < 12.5 \, {\rm GeV}^2$ should be considered for the higher twist  QCD analysis of the non-singlet  data. The value of $\Lambda_{\rm QCD}$ and the free parameters of valence PDFs and  function $h(x)$ could be extracted through the fit procedure to all available data.  
Our strategy is to perform  a fit for the PDF parameters at large values of $W^2$ and  then to impose the TMC effect to the structure functions.
Following that a HT-term is added to the calculation while the PDFs parameters are fixed. The free  parameters of HT-term are determined during the second stage of fit procedure using  the lower values of $W^2$.
Our reason for this strategy of fitting is to indicate the ability of Laplace transformation technique at the NNLO expansion to extract the  non-singlet structure function.  The fitted  result for the
HT term is listed in Table.~\ref{tab:ht}.
\begin{figure}[htb]
	\begin{center}
		\includegraphics[clip,width=0.45\textwidth]{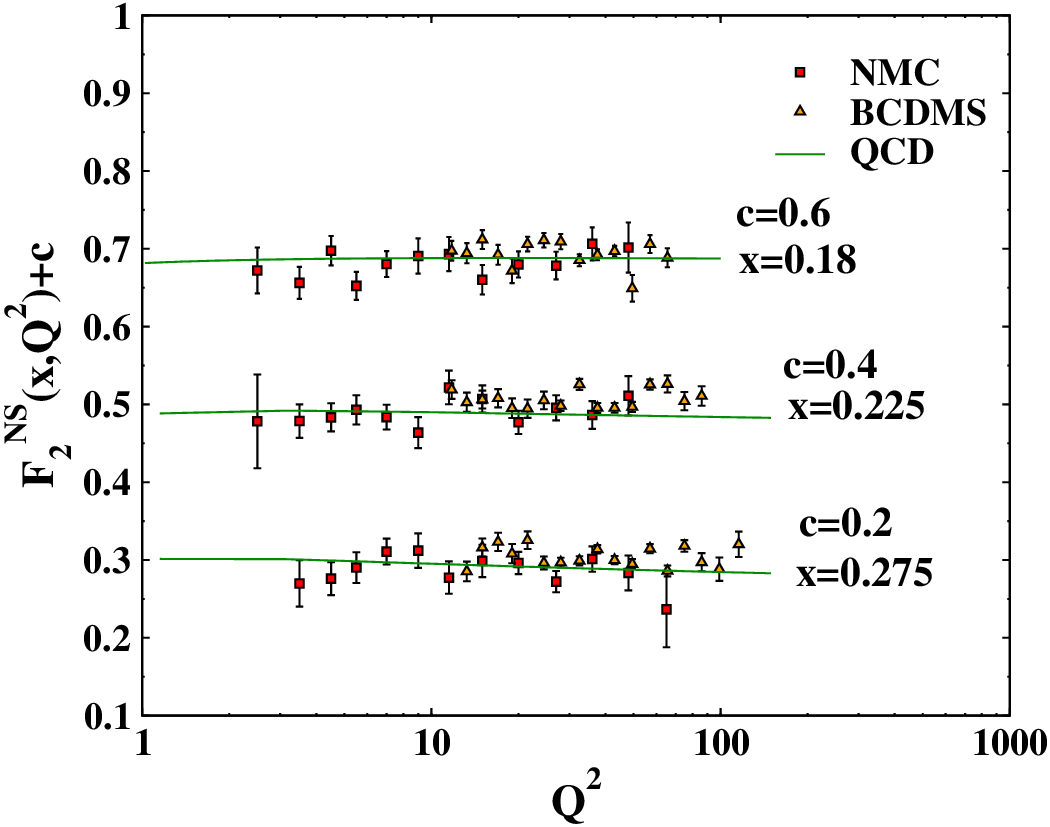}
		\caption{(Color online) Comparison of  our NNLO QCD predictions  for the non-singlet structure function ($F^{\rm NS}_2$) with the dara from {\tt BCDMS} and {\tt NMC}~\cite{Arneodo:1996qe,Arneodo:1995cq}.
		}\label{fig:f2ns}
	\end{center}
\end{figure}
\begin{table}
	\begin{tabular}{c c c c}
		\hline \\\multicolumn{4}{c}{ \rm NNLO}   \\ \hline \hline\\
		$h(x)$	& $ \alpha = 1.07092$ & $ \beta = 0.842737$ & $\gamma = 1.12905$  \\  \hline
	\end{tabular}
	\caption{\label{tab:ht} {\sf The values of HT parameters in Eq.~(\ref{hx}) extracted through the NNLO fit at Q$_0^2 = 2 \, {\rm GeV}^2$. } }
\end{table}

\section{Results and Discussion}
\label{Sec:Results}

In this work, we  discussed about the results arising from the PDF fits. In this regard, we analyzed the global fitting on parton densities in Sec.~\ref{sec:dataset}. They are mostly related to Figs.~\ref{fig:xuvandxdvnnlo}, \ref{fig:partonvalnceQ0} and \ref{fig:deltachinnlo} and also to the corresponding Tables.~\ref{nnlo},  \ref{nlo} and \ref{covmat-matrixNNLO}. In Fig.~\ref{fig:deltachinnlo}, we indicated the results of $\Delta\chi^2$ with respect to the $t$-variable  considering  all five eigenvectors. As can be seen, a very good description of data is provided by considering our theoretical predictions which are based on the analytical solutions resulted from the Laplace transform technique and Jacobi polynomials approach.

The rest of results are related to the TMC and HT effects which have been  described  in Sec.~\ref{Sec:TMC}, in detail. In this section, Fig.~\ref{fig:f2p}  involves the data for inclusive proton  structure functions ($F_2^p$) from {\tt BCDMS} and {\tt SLAC} experiments.  A comparison between available experimental data  and  our NNLO fit, as a function of Q$^2$ at approximately constant values of $x$, has been also done. The effects of TMC
and HT corrections have been  also included so that a considerable agreement between the experimental data and the theoretical predictions is achieved  especially at low energy scales.
Briefly, our strategy includes two steps. In the first step, we performed the required analysis to get the best values for the PDF parameters. At the second step, the TMC and HT effects have been included.  They affected the results, specially at low energy scales. Final results, which are shown in Figs.~\ref{fig:f2p} and \ref{fig:f2d}, indicate  good agreements with the experimental data  for the available range of energy scales.
A more detailed comparison between the theoretical predictions of our NNLO fit  and the data for the deuteron structure function ($F^d_2$)
 has been depicted in Fig.~\ref{fig:f2d}. The results have contained again the effect of TMC and HT  corrections.

In Fig.~\ref{fig:f2ns}, a comparison  between the structure function for $(d-p)$, i.e. $F^{\rm NS}_2$, and  the data from {\tt BCDMS} and
{\tt NMC} experiments has been shown. For better presentation, the data have been scaled by $c=0.2$, $c=0.4$ and $c=0.6$  for $x=0.275$, $x=0.225$ and  $x=0.18$, respectively.

Here, we just remind that the agreement between the theoretical predictions and the experimental data becomes significantly better when one employs the effect of  TMC and HT corrections. This point can be seen clearly  in  Figs.~\ref{fig:f2p} and \ref{fig:f2d}. In this regard, we have also  got  an excellent agreement
between the theoretical prediction for the  structure functions $F^p_2$ and $F^d_2$  and the available experimental
data which included the TMC and HT effects.  The data included  different ranges of Q$^2$ and $x$-values.\\
The technique of Laplace transformation can be also employed in analyzing the Drell-Yan data {{} which contain} antiquark distributions. This type of distribution has not been yet considered  in our recent analysis. It would be also valuable if we can improve the method of our fitting by including the TMC  calculation and the parameters of HT term together with the PDF parameters, simultaneously. These projects are in progress and remain as our future works.

\section*{Acknowledgment}
S.~A.~T is  thankful to the School of Particles and Accelerators, Institute for Research in Fundamental Sciences (IPM) for
financial support of this project. S.~M.~M.~N and A.~M are also grateful  the Yazd university to provide the required facilities to do this project.
\appendix
\section*{Appendix A}
\label{AppendixA}

\begin{widetext}	
In this appendix, we present the Laplace transforms  for the NLO and NNLO splitting functions as well as the Wilson coefficient  functions. They are related to the  non-singlet
sectors of structure function used in Eq.~(\ref{eq:fi-nonsinglet}) and Eqs.~(\ref{eq:Fp})-(\ref{eq:FNS}).
In deriving the NLO and NNLO Wilson coefficients $C_{2NS}^{(1)}$ and $C_{2NS}^{(2)}$  in Laplace $s$-space, we use the corresponding results at $x$-space  given  in Refs.~\cite{Moch:1999eb,Vermaseren:2005qc}.
The NLO and NNLO  splitting functions  in $s$-space, i.e.   $\Phi_{(NS)}^{\rm NLO}$ and $\Phi_{(NS)}^{\rm
NNLO}$,  are calculated  from the corresponding results in $x$-space which are given in  Refs.~\cite{Curci:1980uw,Moch:2004pa}. {{} The obtained results for splitting functions and the Wilson coefficients in $s$-space are as following:}
\begin{eqnarray*}
&&\Phi_{\rm{NS}}^{\rm{NLO}} = \hspace{14 cm}\text{(A.1)}\\
&&C_F T_F\left(-\frac{2}{3 (1+s)^2}-\frac{2}{9 (1+s)}-\frac{2}{3 (2+s)^2}+\frac{22}{9 (2+s)}+\frac{20 \left(\gamma
_E+\psi (s+1)\right)}{9}+\frac{4}{3} \psi '(s+1)\right)+\\
&&C_F{}^2 \left(-\frac{1}{(1+s)^3}-\frac{5}{1+s}-\frac{1}{(2+s)^3}+\frac{2}{(2+s)^2}+\frac{5}{2+s}+\right.\\
&&\frac{2 \left(\gamma _E+\frac{1}{1+s}+\psi (s+1)-(1+s) \psi '(s+2)\right)}{(1+s)^2}+
\frac{2 \left(\gamma _E+\frac{1}{2+s}+\psi (s+2)-(2+s) \psi '(s+3)\right)}{(2+s)^2}-\\
&&\left.4 \left(\left(\gamma _E+\psi (s+1)\right) \psi '(s+1)-\frac{1}{2} \psi ''(s+1)\right)+3 \psi '(s+1)\right)+\\
&&C_A C_F \left(-\frac{1}{(1+s)^3}+\frac{5}{6 (1+s)^2}+\frac{53}{18 (1+s)}+\frac{\pi ^2}{6
(1+s)}-\frac{1}{(2+s)^3}+\right.\\
&&\frac{5}{6 (2+s)^2}-\frac{187}{18 (2+s)}+\frac{\pi ^2}{6 (2+s)}-\frac{67 \left(\gamma _E+\psi (s+1)\right)}{9}+\\
&&\left.\frac{1}{3} \pi ^2 \left(\gamma _E+\psi (s+1)\right)-\frac{11}{3} \psi '(s+1)-\psi ''(s+1)\right)\,,
\end{eqnarray*}

\begin{eqnarray*}
	&& C_{2NS}^{(1)}(s)=\hspace{14 cm}\text{(A.2)} \nonumber\\
	&& C_F \left(-9-\frac{2 \pi ^2}{3}-\frac{2}{(1+s)^2}+\frac{6}{1+s}-\frac{2}{(2+s)^2}+\frac{4}{2+s}+3 \left(\gamma
_E+\psi (s+1)\right)+\right. \nonumber\\
	&& \frac{2 \left(\gamma _E+\psi (s+2)\right)}{1+s}+\frac{2 \left(\gamma _E+\psi (s+3)\right)}{2+s}+ \nonumber\\
	&& \left.\frac{1}{3} \left(\pi ^2+6 \left(\gamma _E+\psi (s+1)\right){}^2-6 \psi '(s+1)\right)+4 \psi
'(s+1)\right)\,,
\end{eqnarray*}
\begin{eqnarray*}
&&\Phi _{\text{NS}}^{\text{NNLO}}\hspace{14 cm}\text{(A.3)}\\
&&=1295.384+\frac{1024}{27
(1+s)^5}-\frac{1600}{9(1+s)^4}+\frac{589.8}{(1+s)^3}-\frac{1258}{(1+s)^2}+\frac{1641.1}{1+s}-\frac{3135}{2+s}+\nonumber\\
&&\frac{243.6}{3+s}-\frac{522.1}{4+s}-1174.898\left(\gamma _E+\psi (1+s)\right)-\frac{714.1\left(\gamma _E+\psi
(2+s)\right)}{1+s}+\nonumber\\
&&\frac{563.9}{(1+s)^2}\left(\gamma _E+\frac{1}{1+s}+\psi (1+s)-(1+s) \psi '(2+s)\right)+\nonumber\\
&&f \left(173.927+\frac{128}{9 (1+s)^4}-\frac{5216}{81
(1+s)^3}+\frac{152.6}{(1+s)^2}-\frac{197}{1+s}+\frac{8.982}{(2+s)^4}+\frac{381.1}{2+s}+\frac{72.94}{3+s}+\right.\nonumber\\
&&\frac{44.79}{4+s}+183.187\left(\gamma _E+\psi (1+s)\right)+\frac{5120 \left(\gamma _E+\psi (2+s)\right)}{81
(1+s)}-\nonumber\\
&&\left.\frac{56.66}{(1+s)^2}\left(\gamma _E+\frac{1}{1+s}+\psi (1+s)-(1+s) \psi '(2+s)\right)\right)-\nonumber\\
&&\frac{256.8 }{(1+s)^4} \left(3+2 \gamma _E (1+s)+2 \gamma _E (1+s)^2 \psi (1+s)-(1+s) \left(-1+2 \gamma _E
(1+s)\right) \psi (1+s)+\right.\nonumber\\
&&(1+s)^3 \psi (1+s)^2 \psi (2+s)-2 (1+s)^3 \psi (1+s) \psi (2+s)^2+(1+s)^3 \psi (2+s)^3-\nonumber\\
&&\left.2 (1+s)^2 \psi '(1+s)+(1+s)^3 \psi \text{''}(2+s)\right)+\nonumber\\
&&f^2 \left(\frac{64 \left(\gamma _E+\psi (1+s)\right)}{81}+\right.\nonumber\\
&&\frac{64}{81} \left(-\frac{51}{16}+\frac{5 \pi ^2}{6}+\frac{3}{2 (1+s)^3}-\frac{11}{2
(1+s)^2}+\frac{7}{1+s}-\frac{3}{2 (2+s)^3}+\frac{11}{2
(2+s)^2}-\frac{6}{2+s}-\right.\nonumber\\
&&\left.\left.3 \zeta (3)-5 \psi '(2+s)- \frac{3}{2} \psi \text{''}(2+s)\right)\right)
\end{eqnarray*}

\begin{eqnarray*}
&&C_{2\text{NS}}^{(2)}=\hspace{14 cm}\text{(A.4)}\nonumber\\
&&-338.513+\frac{160}{9(1+s)^4}-\frac{41.4}{(1+s)^3}+\frac{28.384}{(1+s)^2}-\frac{181}{1+s}+\frac{17.256}{(2+s)^5}-\frac{806.7}{2+s}-\nonumber\\
&&188.641 \left(\gamma _E+\psi (1+s)\right)+\frac{628.8 \left(\gamma _E+\psi (2+s)\right)}{1+s}-\nonumber\\
&&2.5921 \left(\pi ^2+6 \left(\gamma _E+\psi (1+s)\right){}^2-6 \psi '(1+s)\right)+\nonumber\\
&&\frac{72.24 \left(\pi ^2+6 \left(\gamma _E+\psi (2+s)\right){}^2-6 \psi '(2+s)\right)}{6+6 s}+\nonumber\\
&&\frac{24.5166}{(1+s)^2}\left(6 \gamma _E{}^2+\pi ^2+\frac{12 \gamma _E}{1+s}+12 \gamma _E \psi (1+s)-6 (1+s) \psi
(1+s) \psi (2+s)^2+\right.\nonumber\\
&&6 (1+s) \psi (2+s)^3-6 \left(3+2 \gamma _E (1+s)\right) \psi '(2+s)-12 (1+s) \psi (1+s) \psi '(2+s)+\nonumber\\
&&6 (1+s) \psi \text{''}(2+s))-\nonumber\\
&&\frac{37.75}{(1+s)^4}\left(3+2 \gamma _E (1+s)+2 \gamma _E (1+s)^2 \psi (1+s)-(1+s) \left(-1+2 \gamma _E
(1+s)\right) \psi (1+s)+\right.\nonumber\\
&&(1+s)^3 \psi (1+s)^2 \psi (2+s)-2 (1+s)^3 \psi (1+s) \psi (2+s)^2+(1+s)^3 \psi (2+s)^3-\nonumber\allowdisplaybreaks[1]\\
&&\left.2 (1+s)^2 \psi '(1+s)+(1+s)^3 \psi \text{''}(2+s)\right)+\nonumber\\
&&f \left(46.8531+\frac{40}{9 (1+s)^3}-\frac{16}{3
(1+s)^2}-\frac{7.8109}{1+s}+\frac{1.1099}{(2+s)^4}-\frac{17.82}{2+s}-\frac{12.97}{3+s}-\right.\nonumber\\
&&6.34888\left(\gamma _E+\psi (1+s)\right)-\frac{24.87 \left(\gamma _E+\psi (2+s)\right)}{1+s}-\nonumber\\
&&\frac{58}{81} \left(\pi ^2+6 \left(\gamma _E+\psi (1+s)\right){}^2-6 \psi '(1+s)\right)-\nonumber\allowdisplaybreaks[1]\\
&&\frac{15 \left(\pi ^2+6 \left(\gamma _E+\psi (2+s)\right){}^2-6 \psi '(2+s)\right)}{6+6 s}+\nonumber\\
&&\frac{8.113\left(\gamma _E+\frac{1}{1+s}+\psi (1+s)-(1+s) \psi '(2+s)\right)}{(1+s)^2}-\nonumber\\
&&\frac{8}{27} \left(-2 \gamma _E{}^3-\gamma _E \pi ^2-6 \gamma _E \psi (1+s)^2-2 \psi (1+s)^3-\right.\nonumber\\
&&\left.\left.\psi (1+s) \left(6 \gamma _E{}^2+\pi ^2-6 \psi '(1+s)\right)+6 \gamma _E \psi '(1+s)-2 \psi
\text{''}(1+s)-4 \zeta (3)\right)\right)+\nonumber\\
&&\frac{8.87 }{1+s}\left(2 \gamma _E{}^3+\gamma _E \pi ^2+6 \gamma _E \psi (2+s)^2+2 \psi (2+s)^3+\psi (2+s) \left(6
\gamma _E{}^2+\pi ^2-6 \psi
'(2+s)\right)-\right.\nonumber\\
&&\left.6 \gamma _E \psi '(2+s)+2 \psi \text{''}(2+s)+4 \zeta (3)\right)-\nonumber\\
&&\frac{92}{9} \left(-\left(\gamma _E+\psi (1+s)\right) \left(\pi ^2+2 \left(\gamma _E+\psi (1+s)\right){}^2-6 \psi
'(1+s)\right)-\right.\nonumber\\
&&2 (\psi \text{''}(1+s)+2 \zeta (3)))+\nonumber\\
&&\frac{128}{9} \left(\frac{\gamma _E{}^4}{4}+\frac{\gamma _E{}^2 \pi ^2}{4}+\frac{3 \pi ^4}{80}+\gamma _E \psi
(1+s)^3+\frac{1}{4} \psi (1+s)^4+\right.\nonumber\\
&&\frac{1}{4} \psi (1+s)^2 \left(6 \gamma _E{}^2+\pi ^2-6 \psi '(1+s)\right)-\frac{1}{4} \left(6 \gamma _E{}^2+\pi
^2\right) \psi '(1+s)+\nonumber\\
&&\frac{3}{4} (\psi '(1+s))^2+\gamma _E \psi \text{''}(1+s)-\frac{1}{4} \psi \text{'''}(1+s)+2 \gamma _E \zeta
(3)+\nonumber\\
&&\left.\psi (1+s) \left(\gamma _E{}^3+\frac{\gamma _E \pi ^2}{2}-3 \gamma _E \psi '(1+s)+\psi \text{''}(1+s)+2 \zeta
(3)\right)\right)
\end{eqnarray*}
\end{widetext}


%
%

%

\end{document}